\documentclass[a4paper]{article}

\usepackage{amsmath}
\usepackage{amsthm}
\usepackage{amssymb}
\usepackage{amsfonts}
\usepackage[latin1]{inputenc}

\theoremstyle{theorem}

\theoremstyle{definition}

\theoremstyle{remark}

\newcommand{\te}{\theta}

\newcommand\Om\Omega
\newcommand\Te\Theta
\newcommand{\bq}{\mathbf{q}}

\newcommand{\bp}{\mathbf{p}}

\newcommand{\bL}{\mathbf{L}}

\def\RR{\mathbb{R}}

\renewcommand\SS{\mathbb S}
\newcommand{\cD}{{\mathcal D}}

\newcommand{\cM}{{\mathcal M}}

\newcommand\minus\backslash

\newcommand\lan\langle
\newcommand\ran\rangle

\newcommand{\e}{{\mathrm e}}
\newcommand{\dd}{{\mathrm d}}
\newcommand{\sub}[1]{_{\mathrm{#1}}}

\def\bea{\begin{eqnarray}}
\def\eea{\end{eqnarray}}

\def\dd{{\mathrm d}}

\def\1{\'{\i}}

\def\jp{J_+}
\def\jm{J_-}

\def\otra{b}

 \def\te{\theta}

\def\k{\kappa}
\def\>#1{{\mathbf#1}}

\begin{document}

\title{\Large $N$-dimensional $sl(2)$-coalgebra spaces  with non-constant curvature}
\author{\normalsize A. Ballesteros$^a$\thanks{angelb@ubu.es} \and \normalsize
A. Enciso$^b$\thanks{aenciso@fis.ucm.es} \and\normalsize F.J.
Herranz$^a$\thanks{fjherranz@ubu.es} \and\normalsize O.
Ragnisco$^c$\thanks{ragnisco@fis.uniroma3.it}}
\date{\small $^a$ Depto.\ de F\'\i sica, Universidad de Burgos,
09001 Burgos, Spain\vspace{1ex}\\
$^b$ Depto.~de F{\'\i}sica Te\'orica II, 
Universidad Complutense, 
28040 Madrid, Spain\vspace{1ex}\\
$^c$ Dip.\ di Fisica, Universit\`a di Roma 3, and Istituto Nazionale di Fisica
Nucleare,
\\Via Vasca Navale 84, 00146 Rome, Italy}

\maketitle

\begin{abstract}
An infinite  family of $N$D spaces endowed with $sl(2)$-coalgebra symmetry is
introduced. For all these spaces the geodesic flow is superintegrable, and the
explicit form of their common set of integrals is obtained from the underlying
$sl(2)$-coalgebra structure. In particular, $N$D spherically symmetric spaces
with Euclidean signature are shown to be $sl(2)$-coalgebra spaces. As a
byproduct of this construction we present $N$D generalizations of the classical
Darboux surfaces, thus obtaining remarkable superintegrable $N$D spaces with
non-constant curvature.
\end{abstract}

\bigskip




\section{Introduction}

An $N$-dimensional ($N$D) Hamiltonian $H^{(N)}$ is called  completely
integrable if there exists a set of $(N-1)$ globally defined, functionally
independent constants of the motion that Poisson-commute with $H^{(N)}$.
Whereas completely integrable systems are quite unusual~\cite{MM74}, they have
long played a central role in our understanding of dynamical systems and the
analysis of physical models. Moreover, in case that some additional independent
integrals do exist, the system $H^{(N)}$ is called superintegrable~\cite{AKN97}
(there are different degrees of superintegrability, as we shall point out
later). It is well known that superintegrability is strongly related to the
separability of the corresponding Hamilton--Jacobi and Schr\"odinger
equations~\cite{KM84} in more than one coordinate systems, and gives a fighting
chance (which can be made precise in several contexts) of finding the general
solution of the equations of motion by quadratures~\cite{Hu76,Pa81}.

In this paper we consider a specific class of (classical) $N$D Hamiltonian
systems: the geodesic flows on $N$D Riemannian manifolds defined by the
corresponding metrics. Contrary to what happens in the constant curvature
cases, these kinetic-energy Hamiltonians can exhibit extremely complicated
dynamics in arbitrary manifolds, the prime example being the chaotic geodesic
flow on Anosov spaces. The complete integrability of a free Hamiltonian on a
curved space and the separability of its Hamilton--Jacobi equation are rather
nontrivial properties, and the analysis of such systems is being actively pursued
because of its significant connections with the geometry and topology of the
underlying manifold~\cite{KH95,Pa99}.

In physics, curved (pseudo-)Riemannian manifolds (generally, of dimension
higher than four) arise as the natural arena for general relativity,
supergravity and superstring theories, and integrable geodesic flows in
arbitrary dimensions are thus becoming increasingly popular in these
areas~\cite{Gibbons}. Particularly, the case of Kerr--AdS spaces has attracted
much attention due to its wealth of
applications~\cite{Ma00,DMW02,Gauntlett,Cvetic}. In the studies performed so
far, the explicit knowledge of the St\"ackel--Killing integrals of motion in
Kerr--AdS spaces has already proven to be an essential ingredient in these
contexts (see e.g.~\cite{NUT,VasuPRD,Vasu} and references therein), which
suggests that an explicit analysis of integrable geodesic flows on curved
manifolds would certainly meet with interest from this viewpoint. Very
recently, an in-depth analysis of the integrability properties and separability
of the Hamilton--Jacobi equation on Kerr--NUT--AdS spacetimes have been
achieved in~\cite{Kubiznaka, Kubiznakb, Kubiznakc}, thus showing the relevance
of developing the required machinery to deal with superintegrable spaces of
non-constant curvature.

From a quite different perspective, quantum groups (in an $sl_z(2)$ Poisson
coalgebra version) have been recently used to generate a family of
distinguished $N$D  hyperbolic spaces whose curvature is governed by the
deformation parameter $z$ \cite{BHSIGMA}. In these $sl_z(2)$-coalgebra spaces
the geodesic flow is completely integrable and the corresponding $(N-1)$
quadratic first integrals (which give rise to generalized Killing tensors) are
explicitly known. Moreover, these flows turn out to be superintegrable, since
the quantum $sl_z(2)$-coalgebra symmetry provides an additional set of $(N-2)$
integrals. We stress that in several interesting situations (such as in the
$N=2$ case), Lorentzian analogs of these spaces can be obtained through an
analytic continuation method; this procedure has actually been used to
construct a new type of $(1+1)$D integrable deformations of the (Anti-)de
Sitter spaces~\cite{BHRplb}.

In this letter we present a class of $N$D spaces with Euclidean signature whose
geodesic flow is, by construction, superintegrable. This is achieved by making
use of an undeformed Poisson $sl(2)$-coalgebra symmetry. Furthermore, their
$(2N-3)$ constants of the motion, which turn out to be quadratic in the
momenta, are given in closed form. In fact, these invariants have the same form
for all the spaces under consideration as a direct consequence of the
underlying Poisson coalgebra structure, so we can talk about ``universal''
first integrals. As it has been pointed out in \cite{BHletter},
 spaces of constant curvature belong to this class of  $sl(2)$-coalgebra spaces, but
the former are only a small subset of the superintegrable spaces that can be
obtained through this construction. Here we shall present four new significant
$N$D examples with non-constant scalar curvature: the $N$D generalizations of
the so-called (2D) Darboux spaces, which are the only surfaces with
non-constant curvature admitting two functionally independent, quadratic
integrals \cite{Darboux,KKMW02, KKMW03}.

The paper is organized as follows. In the next section we briefly sketch the
construction of generic $sl(2)$-coalgebra spaces and discuss their
superintegrability properties; we also show how spherically symmetric spaces
(with non-constant curvature) arise in this approach. In Section 3 we exploit
the $sl(2)$-coalgebra symmetry of the 2D Darboux spaces to construct $N$D
counterparts. Some brief remarks of global nature are made. Finally, the
closing section includes some comments and open problems.


\section{$sl(2)$-coalgebra spaces and superintegrability}

An $N$D completely integrable Hamiltonian $H^{(N)}$ is called {\it maximally
superintegrable} (MS) if there exists a set of $2N-2$ functionally independent
global first integrals that Poisson-commute with $H^{(N)}$. As is well known,
at least two different subsets of $N-1$ constants in involution can be found
among them. In the same way, a system will be called {\it quasi-maximally
superintegrable} (QMS) if there are $2N-3$ independent integrals with the
aforementioned properties, {\em i.e.} if the system is ``one integral away"
from being MS.

Let us now consider the $sl(2)$ Poisson coalgebra
generated by the following Lie--Poisson brackets and comultiplication map:
\begin{equation}
\begin{array}{l}
\{J_3,J_+\}=2 J_+     ,\quad
\{J_3,J_-\}=-2 J_- ,\quad
\{J_-,J_+\}=4 J_3  ,  \\[4pt]
\Delta(J_l)=  J_l \otimes 1+ 1\otimes J_l ,\qquad l=+,-,3.
\end{array}
 \label{ba}
\end{equation}
The Casimir function is ${\cal C}=  J_- J_+ -J_3^2 $.
Then, the following result holds~\cite{BHletter}:
{Let $\{\>q,\>p \}=\{(q_1,\dots,q_N),(p_1,\dots,p_N)\}$ be $N$ pairs of
canonical variables. The $N$D Hamiltonian
\begin{equation}
{H}^{(N)}= {\cal H}\left(J_-,J_+,J_3\right),
\label{hgen}
\end{equation}
with ${\cal H}$ any smooth function and
\begin{equation}
 J_-=\sum_{i=1}^N q_i^2\equiv \>q^2 ,\  \    J_+=
    \sum_{i=1}^N \left(  p_i^2+\frac{\otra_i}{ q_i^2} \right)
\equiv \>p^2 +  \sum_{i=1}^N  \frac{\otra_i}{ q_i^2}  ,\  \ J_3=
  \sum_{i=1}^N  q_i p_i\equiv \>q\cdot\>p  ,
\label{qp}
\end{equation}
where $b_i$ are arbitrary real parameters, is a QMS system. The   $(2N-3)$
functionally independent ``universal" integrals of motion for ${H}^{(N)}$ read
\bea && \!\!\!\!\!\!\!\!\!\!\!  C^{(m)}= \sum_{1\leq i<j}^m \left\{ ({q_i}{p_j}
- {q_j}{p_i})^2 + \left(
\otra_i\frac{q_j^2}{q_i^2}+\otra_j\frac{q_i^2}{q_j^2}\right)\right\}
+\sum_{i=1}^m \otra_i , \nonumber\\
&& \!\!\!\!\!\!\!\!\!\!\!  C_{(m)}= \!\!\sum_{N-m+1\leq i<j}^N \left\{ ({q_i}{p_j} -
{q_j}{p_i})^2 + \left(
\otra_i\frac{q_j^2}{q_i^2}+\otra_j\frac{q_i^2}{q_j^2}\right)\right\}
+\!\! \sum_{i=N-m+1}^N \!\!\!\!\otra_i  ,
\label{rightc}
\eea
where $m=2,\dots, N$ and $C^{(N)}=C_{(N)}$.  Moreover, the sets of $N$ functions
$\{H^{(N)},C^{(m)}\}$ and $\{H^{(N)},C_{(m)}\}$ $(m=2,\dots, N)$ are in
involution.
}

The proof of this result is based on the fact that, for any choice of the
function ${\cal H}$, the Hamiltonian $H^{(N)}$ has an $sl(2)$ Poisson coalgebra
symmetry~\cite{BHSIGMA,BHletter}; the generators (\ref{qp}) fulfil the
Lie--Poisson brackets of $sl(2)$ and the integrals (\ref{rightc}) are obtained
through the $m$-th coproducts of the Casimir $\cal C$ within an $m$-particle
symplectic realization of type (\ref{qp}).

With the previous general result in mind, we shall say that an $N$D Riemannian
manifold is an $sl(2)$-coalgebra space if the kinetic energy Hamiltonian
$H^{(N)}_T$ corresponding to geodesic motion on such a space has
$sl(2)$-coalgebra symmetry, i.e., if  $H^{(N)}_T$ can be written as
\begin{equation}\label{HT}
{H}^{(N)}_T= {\cal H}_T\left(J_-,J_+,J_3\right)={\cal H}_T\left( \>q^2, \>p^2+
\sum_{i=1}^N  \frac{\otra_i}{ q_i^2} , \>q\cdot\>p\right ),
\end{equation}
where ${\cal H}_T$ is some smooth function on the $sl(2)$-coalgebra
generators~\eqref{qp}. Since ${H}^{(N)}_T$ has to be homogeneous quadratic in
the momenta, we are forced to restrict ourselves to the specific $sl(2)$
symplectic realizations~\eqref{qp} with all $\otra_i=0$, so that the most
general Hamiltonian corresponding to an $sl(2)$-coalgebra space reads
\begin{equation}\label{HTgeneric}
{\cal H}_T=  {\cal A}(\jm)\, \jp + {\cal B}(\jm)\, J_3^2
= {\cal A}({\>q}^2)\, {\>p}^2 + {\cal B}({\>q}^2)\, (\>q\cdot\>p)^2,
\end{equation}
where ${\cal A}$ and ${\cal B}$ are arbitrary functions. At this point, we
stress that for {any} choice of both functions, ${H}^{(N)}_T$ is a QMS
Hamiltonian system with integrals given by (\ref{rightc}). Thus, an infinite
family of $N$D spaces with QMS geodesic flow is defined by (\ref{HTgeneric})
or, equivalently, by the pair of functions $({\cal A},{\cal B})$ that will
characterize the $N$D metric.


\subsection {Spaces of constant curvature}

In the previous discussion it is implicit that the pair $(\>q,\>p)$ is an
arbitrary set of canonically conjugated positions and momenta, for which no
{\it a priori} geometric interpretation is given. This becomes apparent by
considering the (simply connected) $N$D Riemannian spaces with constant
sectional curvature $\kappa$ (the sphere  $\SS^N$ and the hyperbolic $\mathbb H^N$ space), which are
distinguished examples of $sl(2)$-coalgebra spaces. It can be shown
\cite{BHletter} that the corresponding Hamiltonians can be written in the
following ways (among others), both of them compatible with (\ref{HTgeneric}):
\begin{equation}
\begin{array}{l}
\displaystyle{ {\cal H}^{\rm P}_T=\frac{1}{2}\left( 1+\k
J_-\right)^2 J_+=
\frac{1}{2}\left( 1+\k \>q^2\right)^2 \>p^2} ,\\[8pt]
\displaystyle{ {\cal H}^{\rm B}_T=\frac{1}{2}\left( 1+\k
J_-\right)\left(  J_+ +\k J_3^2\right)=
\frac{1}{2}(1+\k \>q^2)\left( \>p^2+\k (\>q\cdot \>p)^2 \right) }.
\end{array}
\label{dd}
\end{equation}
The associated coordinate systems are classical: in the first case, $\>q$
denotes the Poincar\'e coordinates in $\SS^N$ or $\mathbb H^N$, coming from the
stereographic projection in ${\mathbb R^{N+1}}$~\cite{Doub}, whereas in the
second one $\>q$ are the Beltrami coordinates, which are associated with the
central projection. We recall that the image of both the  stereographic
projection (Poincar\'e coordinates) and  central projection (Beltrami
coordinates)  is the subset of   $\ \mathbb  R^N$   determined by $  1+\k
\>q^2>0$, which means that for   $\mathbb H^N$ with $\k=-1$    such an image is
the open subset $\>q^2<1$. In both cases, we recover the standard Cartesian
coordinates in $\mathbb  R^N$ when we set $\k=0$.

Note that in this language both Hamiltonians can immediately be interpreted as
deformations (in terms of the curvature parameter $\k$) of the motion on
Euclidean space $\mathbb E^N$, to which they reduce when $\k=0$. This is
analogous to the analysis in terms of the quantum parameter $z$ carried out in
\cite{BHRplb}. By construction, the above Hamiltonians admit the universal
integrals (\ref{rightc}), whose concrete geometric realization depends on the
interpretation of $(\>q,\>p)$ as either Poincar\'e or Beltrami coordinates. The
generalized Killing vectors of these spaces are   the Hamiltonian vector fields
(in phase space) associated with the latter first integrals. It should be noted
that these spaces admit in fact an additional first integral which makes them
MS: while they are not strictly the only ones having this property
(see~\cite{BEHR07} and references therein), this final symmetry is not of
coalgebraic nature and, when it exists, must be found by {\em ad hoc} methods.


\subsection {Spherically symmetric spaces}

Any $N$D spherically symmetric metric of the type
\begin{equation}\label{metric}
\dd s^2=f(|\bq|)^2\,\dd\bq^2 ,
\end{equation}
where $|\bq|=\sqrt{\bq^2}$,  $\dd\bq^2 =\sum_{i}\dd q_i^2$ and $f$ is an
arbitrary smooth function, leads to a geodesic motion described by the
Hamiltonian
\begin{equation}\label{H0}
H_T^{(N)}=\frac{\bp^2}{f(|\bq|)^2}\,,
\end{equation}
which is clearly of the form (\ref{HTgeneric}). Therefore, the metric (\ref{metric})
  corresponds to   an $sl(2)$-coalgebra space with
\begin{equation}\label{Af}
{\cal A}(\jm)=\frac{1}{f(\sqrt{\jm})^2},\quad {\cal B}(\jm)=0,
\end{equation}
so that for any choice of $f$ its geodesic flow defines a QMS system whose
generalized Killing symmetries are the Hamiltonian vector fields associated
with (\ref{rightc}). It is apparent that these spaces are conformally flat, so
that its Weyl tensor vanishes. The scalar curvature $R$ of (\ref{metric}),
which is generally non-constant, can be computed to be
\begin{equation}\label{curv}
R=-(N-1)\left( \frac{    (N-4)f'(|\bq|)^2+  f(|\bq|)  \left(    2f''(|\bq|)+2(N-1)|\bq|^{-1}f'(|\bq|)  \right)}   {f(|\bq|)^4  } \right) .
\end{equation}

If we define $N$D spherical coordinates $(r,\te_1,\dots,\te_{N-1})$ as
$$
q_j=r \cos\te_{j}     \prod_{k=1}^{j-1}\sin\te_k , \quad
 q_N =r \prod_{k=1}^{N-1}\sin\te_k  ,
\label{bb}
$$
where $j=1,\dots,N-1$, $r=|\bq|$ and hereafter a product $\prod_{k=1}^0$   is assumed to be equal to 1, the metric \eqref{metric} can be
alternatively written as
\[
\dd s^2=f(r)^2(\dd r^2+r^2\dd\Om^2_{N-1})\,.
\]
Here
\[
\dd\Om^2_{N-1}=\sum_{j=1}^{N-1}\dd\te_j^2\prod_{k=1}^{j-1}\sin^2\te_k,
\]
denotes the metric of the unit $(N-1)$-sphere $\SS^{N-1}$, with
$\dd\Om_1^2=\dd\te_1^2$. In these coordinates the free Hamiltonian~\eqref{H0}
can be equivalently expressed as
\[
H_T^{(N)}=\frac{p_r^2+r^{-2}\bL^2}{f(r)^2}\,,
\]
where
\begin{equation}\label{L2}
\bL^2=\sum_{j=1}^{N-1}p_{\te_j}^2\prod_{k=1}^{j-1}(\sin\te_k)^{-2},
\end{equation}
is the squared angular momentum and $(p_r,p_{\te_1},\dots,p_{\te_{N-1}})$ are
the conjugate momenta of $(r,\te_1,\dots,\te_{N-1})$.

Finally, it is also convenient for our purposes to consider the modified
spherical system given by $(\rho,\te_1,\dots,\te_{N-1})$, where $\rho=\ln r$.
If $p_\rho$ stands for the conjugate momentum of $\rho$, this yields
\begin{align}
\dd s^2&=F(\rho)^2\,(\dd \rho^2+\dd\Om_{N-1}^2)\,,\label{dsrho}\\
H_T^{(N)}&=\frac{p_\rho^2+\bL^2}{F(\rho)^2}\label{H0rho}\,,
\end{align}
where the arbitrary function $F$ is defined as $F(\rho)=r\,f(r)$.
Hence any metric of the form (\ref{dsrho})  defines an $sl(2)$-coalgebra
space given by
\begin{equation}\label{AFuv}
{\cal A}(\jm)=\frac{\jm}{F(\ln \sqrt{\jm})^2},\quad {\cal B}(\jm)=0,
\end{equation}
and the geodesic flows on (\ref{dsrho}) are QMS for any choice of $F$. The scalar curvature (\ref{curv}) now reads
\begin{equation}\label{curvrho}
R=-(N-1)\left( \frac{    (N-4)F'(\rho)^2-  (N-2)F(\rho)^2 +2  F(\rho) F''(\rho)  }   {F(\rho)^4    } \right) .
\end{equation}


\section{Darboux spaces}

The (2D) Darboux surfaces are the 2-manifolds with non-constant curvature
admitting two quadratic first integrals, so that its geodesic motion is
quadratically MS. There are only four types of such spaces~\cite{Darboux},
which we will represent by $\cD_i^{(2)}$ ($i=\mathrm{I,II,III,IV}$) following
the notation in~\cite{KKMW02,KKMW03}. In this section, the spaces $\cD_i^{(2)}$
will be initially described in terms of isothermal coordinates~\cite{Do76}
$(u,v)$ with canonically conjugate momenta $(p_u,p_v)$. In~\cite{KKMW02,KKMW03}
it has been explicitly shown that their natural free Hamiltonians can be
expressed in these variables as
\begin{equation}
H^{(2)}=\frac{p_u^2+p_v^2}{F(u)^2}\,,
\end{equation}
which implies that
\begin{equation}\label{ds2}
\dd s^2=F(u)^2\,(\dd u^2+\dd v^2).
\end{equation}
Occasionally we shall need to consider other different isothermal charts of
$\cD^{(2)}_i$ with isothermal coordinates  $(\xi,\eta)$ and conjugate momenta
$(p_\xi,p_\eta)$.

Immediately, from \eqref{ds2} we realize that the  four 2D Darboux spaces
$\cD_i^{(2)}$ are  $sl(2)$-coalgebra spaces of the type $({\cal A},0)$ with
${\cal A}(\jm)$ given by \eqref{AFuv}. As a consequence, we can use the
underlying $sl(2)$-coalgebra symmetry to define $N$D, spherically symmetric,
conformally flat generalizations $\cD_i^{(N)}$ of the Darboux surfaces that
will be thoroughly described in the following subsections. By construction, the
four $N$D spaces so constructed will have QMS geodesic motions and their
$(2N-3)$ independent integrals will be given by  (\ref{rightc}). It should be
highlighted that, at least for the space $\cD_{\rm III}^{(N)}$ the additional
integral giving rise to an $N$D MS system can  be explicitly
constructed~\cite{BEHR07}.


\subsection{Type I}

The Hamiltonian for geodesic motion on $\cD_{\rm I}^{(2)}$is given by
\[
H_{\rm I}^{(2)}=\frac{p_u^2+p_v^2}u .
\]
Therefore the   corresponding metric
reads~\cite{KKMW02}
\begin{equation}\label{ds2I}
\dd s^2 =u(\dd u^2+\dd v^2)\,.
\end{equation}
The construction of the $N$D space $\cD_{\rm I}^{(N)}$
 can be conveniently performed via the substitution
\begin{equation}\label{subst}
u\to \rho=\ln r\,,\qquad\dd v^2\to\dd\Om^2_{N-1}\,.
\end{equation}
Note that we have used the same letter for the functions $F(u)$ and $F(\rho)$
appearing in the coordinate expressions of the metric (Eqs.~\eqref{ds2}
and~\eqref{dsrho}) because upon this substitution they define, in fact, the
same function $F:\RR\to\RR^+$.

Therefore, we find that the generic $N$D metric~\eqref{dsrho} is in this case
characterized by the function $F(\rho)=\rho^{1/2}$, yielding   the
following metric  for   $\cD_{\rm I}^{(N)}$:
\begin{equation}\label{dsI}
\dd s^2 =\rho\,(\dd \rho^2+\dd\Om_{N-1}^2)=\frac{\ln|\bq|\,\dd\bq^2}{\bq^2}.
\end{equation}
In other words, $\cD_{\rm I}^{(N)}$ is the $sl(2)$-coalgebra space $({\cal A},0)$  given by
\begin{equation}\label{AFDI}
{\cal A} (\jm)=\frac{\jm}{\ln\sqrt{\jm}}.
\end{equation}
This space is certainly not flat; its scalar curvature  can be readily
computed to be
\begin{align*}
R &=\frac{(N-1) \left(4 (N-2) \rho^2-N+6\right)}{4 \rho^3}\,.
\end{align*}

Some remarks on the global properties of the type I Darboux $N$-manifold are in
order. We define $\cD\sub I^{(N)}$ to be the exterior of the closed unit ball
\[
\cM_+=\big\{\bq:|\bq|>1\big\}\,,
\]
covered with the coordinates $\bq$ and endowed with the metric~\eqref{dsI}. It
is not difficult to see that this space is incomplete by integrating its radial
geodesics. In fact, the radial motion on $\cD\sub I^{(N)}$ is obtained from the
Lagrangian $L=r^{-2}\ln r\,\dot r^2\equiv G(r)^2\,\dot r^2$, so that a
straightforward calculation shows that the radial geodesics are complete at
infinity and incomplete at the hypersphere $|\bq|=1$ since the integral $\int
G(r)\,\dd r$ diverges at infinity but converges at $1$. It should be
remarked  that one can also replace the conformal factor $\ln r$ by its
absolute value and define $\cD\sub I^{(N)}$ to be the interior of the unit ball
\[
\cM_-=\big\{\bq:|\bq|<1\big\}
\]
together with the metric~\eqref{dsI}. The latter manifold is complete at 0 and
incomplete at 1.


\subsection{Type II}

In this case the free Hamiltonian reads~\cite{KKMW03}
\[
H_{\rm II}^{(2)}=\frac{p_u^2+p_v^2}{1+u^{-2}}\,.
\]
The QMS $N$D extension  is performed again using the
substitution~\eqref{subst}. In this case, $F(\rho)=(1+\rho^{-2})^{1/2}$ and we have
\begin{align*}
H_{\rm II}^{(N)}=\frac{p_\rho^2+\bL^2}{1+\rho^{-2}}
=\frac{\bq^2}{1+(\ln|\bq|)^{-2}}\,\bp^2\,.
\end{align*}
Thus, $\cD_{\rm II}^{(N)}$ is the $sl(2)$-coalgebra space determined by
\begin{equation}\label{AFDII}
{\cal A} (\jm)=\frac{\jm}{1+(\ln\sqrt{\jm})^{-2}}.
\end{equation}
The metric of $\cD\sub{\rm II}^{(N)}$ is then given by
\[
\dd s^2=(1+\rho^{-2})  (\dd
\rho^2+\dd\Om_{N-1}^2)=\frac{1+(\ln|\bq|)^{-2}}{\bq^2}\, \dd\bq^2\,,
\]
with non-constant scalar curvature
\begin{align*}
R=\frac{(N-1) \big[N [(\rho ^3+\rho )^2-1]-2 \rho ^2 (\rho ^4+2 \rho
^2+4)\big]}{(\rho ^2+1)^3}\,.
\end{align*}

If we set $G(r)^2=r^{-2}(1+\ln^{-2}r)$, the same arguments discussed in the
previous subsection show that the radial geodesics of $\cD\sub{II}^{(N)}$ are
complete at 0, 1 and at infinity. Hence both manifolds $(\cM_\pm,\dd s^2)$ are
complete.


\subsection{Type III}

The 2D  free Hamiltonian reads now~\cite{KKMW03}
\[
H_{\rm III}^{(2)}=\frac{\e^{2u}}{1+\e^u}(p_u^2+p_v^2)\,.
\]
In order to obtain the $N$D spherically symmetric generalization, it suffices
to take $F(\rho)=\e^{-\rho}(1+\e^\rho)^{1/2}$ and apply the map (\ref{subst}),
so that the metric of $\cD\sub{\rm III}^{(N)}$ becomes
\[
\dd s^2=  \e^{-2\rho}(1+\e^\rho) (\dd
\rho^2+\dd\Om_{N-1}^2)=\        \frac{1+|\bq|}{\bq^4}\,\dd\bq^2\,,
\]
and the space is characterized by
\begin{equation}\label{AFDIII}
{\cal A}(\jm)=\frac{\jm^2}{1+\sqrt{\jm}}.
\end{equation}
The scalar curvature reads (recall that $r={\rm e}^\rho$):
\begin{align*}
R=\frac{r^3 (N-1) [N (3 r+4)-6 (r+2)]}{4 (r+1)^3}\,.
\end{align*}

In this case the radial geodesics are obtained from the Lagrangian
$L=r^{-4}(1+r)\,\dot r^2\equiv G(r)^2\,\dot r^2$. The integral $\int G(r)\,\dd
r$ diverges at 0 but is finite at $\infty$, and therefore $\cD\sub{\rm
III}^{(N)}=(\RR^{(N)}\backslash\{\mathbf0\},\dd s^2)$ is complete at $\mathbf0$ but
incomplete at $\infty$ (i.e., free particles in this space escape to infinity
in finite time). Again, the whole manifold is conveniently described in terms
of the coordinates $\bq$.

It should be pointed out that the Hamiltonian $H_{\rm III}^{(2)}$ can be
written in a different coordinate system $(\xi,\eta)$ as
\[
H_{\rm III}^{(2)}=\frac{p_\xi^2+p_\eta^2}{1+\xi^2+\eta^2}\,,
\]
which admits the $N$D coalgebraic generalization
\[
H_{\rm III}^{(N)}=\frac{\bp^2}{1+\bq^2}=\frac{J_+}{1+J_-}\,.
\]
The complete manifold $(\RR^N\backslash\{\mathbf0\},(1+\bq^2)\,\dd\bq^2)$ was
thoroughly studied in~\cite{BEHR07}, showing that it is in fact MS.


\subsection{Type IV}

The Darboux Hamiltonian of type IV is  given by~\cite{KKMW03}
\begin{equation}
H_{\rm IV}^{(2)}=\frac{\sin^2u}{a+\cos u}\Big(p_u^2+p_v^2\Big),
\end{equation}
where $a$ is a constant. This Hamiltonian admits a QMS  $N$D generalization    via the
substitution
\[
u\to\rho=\ln r\,,\qquad p_v^2\to\bL^2 ,
\]
with $F(\rho)=\sin^{-1}\rho\,(a+\cos\rho)^{1/2}$. More precisely, the system
has the form
\[
H_{\rm
IV}^{(N)}=\frac{\sin^2\rho}{a+\cos\rho}\Big(p_\rho^2+\bL^2\Big)=\frac{\bq^2\sin^2(\ln|\bq|)}{a+\cos(\ln|\bq|)}\,\bp^2\,,
\]
so that the $sl(2)$-coalgebra space corresponds to setting
\begin{equation}\label{AFDIV}
{\cal A}(\jm)=\frac{\jm\sin^2(\frac12\ln{\jm})}{a+\cos(\frac12\ln{\jm})},
\end{equation}
and
the metric in $\cD\sub{\rm IV}^N$ is given by
\[
\dd s^2=\frac{a+\cos\rho}{\sin^2\rho}\,(\dd
\rho^2+\dd\Om_{N-1}^2)=
\frac{a+\cos(\ln|\bq|)}{\bq^2\sin^2(\ln|\bq|)}\,\dd\bq^2\, .
\]
Its scalar curvature   is found to be
\begin{multline*}
R=-\frac{N-1}{32 (a+\cos (\rho ))^3}\Big(64 a^2+40 (N+1) \cos (\rho ) a+8 (3
N-5) \cos (3 \rho ) a+15 N\\
+4[8 (N-2) a^2+3 (N+2)] \cos (2 \rho )+5 (N-2) \cos (4 \rho)-14\Big)\,.
\end{multline*}

If we take $a>1$, it is obvious from inspection that the metric becomes
singular at $r=1$ and $r=\e^\pi$. The radial geodesics are obtained from the
Lagrangian $L=\sin^{-2}\rho\,(a+\cos\rho)\,\dot\rho^2\equiv
G(\rho)^2\,\dot\rho^2$. As the integral $\int G(\rho)\,\dd\rho$ diverges both at
0 and $\pi$, it immediately follows that the Riemannian manifold $\cD\sub{\rm
IV}^{(N)}=(M,\dd s^2)$ is complete, $M$ being the annulus
\[
M=\big\{\bq\in\RR^N:1<|\bq|<\e^\pi\big\}\,.
\]


\section{Concluding remarks}

In the framework here discussed, the  notion of $sl(2)$-coalgebra spaces arises
naturally when analyzing (generalized) symmetries in Riemannian manifolds, and
can be rephrased in terms of an $sl(2)\otimes sl(2)\otimes \cdots^{(N)}\otimes
sl(2)$ dynamical symmetry of the free Hamiltonian on these spaces. As a matter
of fact, we have shown that spherically symmetric spaces are $sl(2)$-coalgebra
ones. We stress that once a non-constant curvature space is identified within
the family (\ref{HTgeneric}), the underlying coalgebra symmetry ensures that
this is, by construction, QMS. It should be explicitly mentioned that not every
integrable geodesic flow is amenable to the $sl(2)$-coalgebraic approach
developed in this paper by means of an appropriate change of variables. For
instance, the completely integrable Kerr--NUT--AdS spacetime studied
in~\cite{Kubiznaka} does not fit within this framework, even after
euclideanization.

Moreover, any potential with $sl(2)$-coalgebra symmetry, {\em i.e.}\ given by a
function $V\left(J_-,J_+,J_3\right)$, can be added to the kinetic energy $H_T$
of an $sl(2)$-coalgebra background space without breaking the
superintegrability of the motion. In this respect, we stress that the
symplectic realization (\ref{qp}) with arbitrary parameters $\otra_i$'s would
give rise to potential terms of ``centrifugal" type. It is well known that the
latter terms can be often added to some ``basic" potentials (such as the
Kepler--Coulomb and the harmonic oscillator potentials) without breaking their
superintegrability.

Among the infinite family of  $sl(2)$-coalgebra spaces, the four $N$D Darboux
spaces here introduced are from an algebraic viewpoint the closest ones to
constant curvature spaces, since they are the only spaces other than $\mathbb
E^N$, $\mathbb H^N$ and $\SS^N$ whose geodesic motion can be expected to be
(quadratically) MS for all $N$. In the case $N=2$, this statement is the
cornerstone of Koenigs classification \cite{Darboux}, whereas in the case of
$\cD\sub{\rm III}^{(N)}$ such maximal superintegrability has been recently
proven in \cite{BEHR07}. The search for the additional independent integral of
motion in the three remaining Darboux spaces is currently under investigation,
as is the exhaustive analysis of $N$D versions of the 2D
potentials given in \cite{KKMW02,KKMW03}.

Another interesting problem is the construction of the Lorentzian counterparts
of the Riemannian $sl(2)$-coalgebra spaces presented in this letter. We expect
that such an extension should be feasible by resorting to an analytic
continuation procedure similar to the one used in \cite{BHRplb}. In this
direction, we believe that an appropriate shift to the Lorentzian signature
should not affect the superintegrability properties of the geodesic flows, in
the same way that separability is not altered by the standard analytic
continuation tecniques~\cite{Vasu}.


\section*{Acknowledgements}

This work was partially supported by the Spanish MEC and by the Junta
de Castilla y Le\'on under grants no.\ FIS2004-07913 and VA013C05
(A.B.\ and F.J.H.), by the Spanish DGI under grant no.\
FIS2005-00752 (A.E.) and by the INFN--CICyT (O.R.). Furthermore,
A.E. acknowledges the financial support of the Spanish MEC through
an FPU scholarship, as well as the hospitality and the partial support of the Physics Department of Roma Tre University.


\end{document}